\documentclass[12pt]{iopart}
\usepackage{iopams,amssymb,hyperref,graphicx,subfigure}

\begin{document}
\title{Competition for finite resources}

\author{L. Jonathan Cook$^1$ and R. K. P. Zia$^2$}

\begin{abstract}
The resources in a cell are finite, which implies that the various components of the cell must compete for resources.  One such resource is the ribosomes used during translation to create proteins.  Motivated by this example, we explore this competition by connecting two totally asymmetric simple exclusion processes (TASEPs) to a finite pool of particles.  Expanding on our previous work, we focus on the effects on the density and current of having different entry and exit rates.
\end{abstract}

\address{1.  Department of Physics and Engineering, Washington \& Lee University, Lexington, VA 24550, USA\\
2.  Department of Physics, Virginia Tech, Blacksburg, VA 24061-0435 and 
Department of Physics and Astronomy, Iowa State University, Ames, IA 50011-3160, USA}

\ead{cookj@wlu.edu, rkpzia@vt.edu}


\section{Introduction}

Non-equilibrium systems are ubiquitous in nature. With no comprehensive
framework for such systems in general, the understanding of non-equilibrium
statistical mechanics is recognized as one of the major challenges \cite%
{CMMP10}. To make progress toward finding such a framework, it is reasonable
to study simplified models, in order to gain some insight into this type
of complex systems. One such model is totally asymmetric simple exclusion
process (TASEP). On the one hand, this model is simple enough to be amenable
to analytic methods, so that many exact results are known. At the same time,
it is applicable to a wide range of biological and
physical systems, e.g. protein production \cite{MacDonald68, MacDonald69,
Shaw03,Chou03}, traffic flow \cite{Chowdhury00, Popkov01}, and surface growth \cite%
{Kardar86, Wolf90}.

The simplest version of the TASEP consists of a one-dimensional lattice with
particles moving unidirectionally from one site to the next. Particles may
move only if the adjacent site is empty. Two types of boundary conditions are
typically studied, periodic and open. With periodic boundary conditions, the
stationary distribution is trivial \cite{Spitzer70}, though its dynamics
differ from that of ordinary diffusion \cite{DeMasi85, Kutner85, Dhar87,
Majumdar91, Gwa92, Derrida93b, Kim95, Golinelli05}. For open
boundary conditions, three distinct phases emerge that depend on the entry
and exit rates \cite{Krug91} - a low density (LD) phase with the lattice
less than half filled, a high density (HD) phase with more than half of the
lattice filled, and a maximal current (MC) phase where the current of
particles through the lattice is a maximum. If the entry and exit rates are
the same, then a shock forms between a LD and HD region that performs a
random walk on the lattice. Because of the presence of a shock, it is often
referred to as the shock phase (SP). The exact solution of the steady-state
distribution is non-trivial and was found only two decades ago \cite%
{Derrida92, Derrida93, Schutz93}. Not surprisingly, its dynamics is more
complex \cite{Pierobon05, Dudzinski00, Nagy02, Takesue03, deGier06, Gupta07}. For a recent review on these aspects of the TASEP, as well as its applications to other processes of biological transport, see \cite{Chou11}.

The TASEP with open boundary conditions has been used to study the
production of proteins during translation in a cell \cite{MacDonald68,
MacDonald69, Shaw03}. In this process, ribosomes attach at one end of the
messenger RNA (mRNA) strand and move unidirectionally to the other end. At
the other end, the ribosome detaches from the strand and can be used again
by either the same mRNA or another one. To build more realistic models for
protein synthesis, modifications to the simplest version of the TASEP have
been introduced, such as having large particles \cite{Chou03,Dong07},
inhomogeneous hopping rates \cite{Chou04,Dong07b}, and ribosome
\textquotedblleft recycling\textquotedblright\ \cite{Chou03b, Adams08,
Cook09}. In this paper, we expand on our previous work on competition
between multiple TASEPs\cite{Cook09b}, modeling the simultaneous translation
of multiple genes in a cell with a limited number of ribosomes. Unlike
earlier studies, we consider another important aspect of synthesis of
proteins in a cell, i.e., the presence of various regulatory mechanisms
which control the rates of ribosome binding to different proteins. Thus, we
study TASEP's with {\em different} entry and exit rates. Though we are not
aware of any similar mechanism for termination, we consider different exit
rates also, simply as part of a systematic investigation. With such a large
parameter space to explore, we restrict ourselves to only two TASEPs here,
in search for novel and (possibly) universal properties that could be
applicable for mRNA competition in a real cell.

This paper is organized as follows:  In the next section, we define our model. In section \ref{Section3}, we present our simulation results. We give some theoretical considerations in section \ref{Section4}. Finally, we give a summary and outlook in section \ref{Section5}.

\section{Model specifications}

In our previous study \cite{Cook09b}, we model the competition between mRNAs
by coupling two or more open TASEPs to a finite pool of $N_{p}$ particles
and let the entry rates depend on this $N_{p}$. Particles exiting each
TASEP join this pool and are \textquotedblleft recycled\textquotedblright\
for entry into any of the other TASEPs. Thus, the total number of particles $%
N_{tot}$ is conserved. While on any lattice, the particles move
uni-directionally from one side to the other as in the ordinary TASEP. All
internal hopping rates are set to unity. 
\begin{figure}[htb]
\begin{center}
\includegraphics[width=0.5\textwidth]{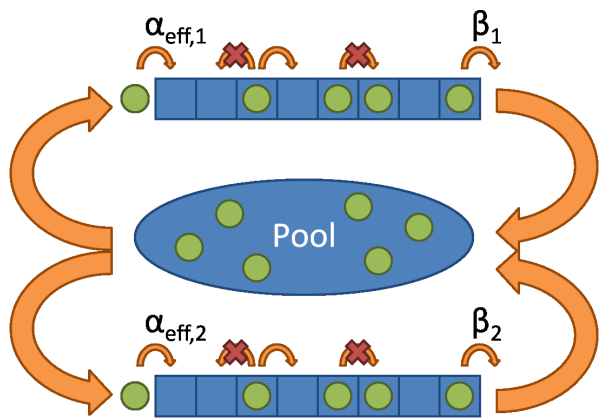}
\caption{Our current model of connecting two TASEPs to a finite pool of particles.  The large arrows indicate how particles enter and leave the pool.}
\label{model}
\end{center}
\end{figure}
Our current model (shown in figure \ref{model}) differs from \cite%
{Cook09b}: Here, we relax the constraint that the {\em intrinsic }(i.e.,
limiting) entry rates of the TASEPs are identical. Thus, we define $\alpha
_{1,2}$ as the intrinsic rate for our two-TASEP system, applicable when the
supply of particles is very large. For simplicity, let us assume the
crossover function ($f$) to be the same, so that the {\em effective }entry
rates are given by 
\begin{eqnarray}
\alpha _{eff,1}& =\alpha _{1}f(N_{p})  \label{a-eff1} \\
\alpha _{eff,2}& =\alpha _{2}f(N_{p})  \label{a-eff2}
\end{eqnarray}%
As in \cite{Adams08, Cook09, Cook09b}, we will use 
\begin{equation}
f(N_{p})=\tanh \left( \frac{N_{p}}{N^{\ast }}\right)  \label{f-def}
\end{equation}%
(where $N^{\ast }$ is a scaling parameter), so that $f\left( 0\right) =0$
and $f\rightarrow 1$ as $N_{p}\rightarrow \infty $. Clearly, it is
reasonable to use the labels \textquotedblleft faster\textquotedblright
/\textquotedblleft slower\textquotedblright\ TASEP for the one with
larger/smaller $\alpha $. We also consider different exit
rates $\beta _{1,2}$, even though we are not aware of biological systems
which exhibit such differences.

In our Monte Carlo simulations, we first consider the case of two TASEPs of
lengths $L_{1}$ and $L_{2}$ connected to a single pool of particles.To
represent the pool, we have a \textquotedblleft virtual\textquotedblright\
site, with unlimited occupation (so that we have $L_{1}+L_{2}+1$ sites in
total). Since this site is connected to both TASEPs, there are actually $%
L_{1}+L_{2}+2$ \textquotedblleft bonds\textquotedblright\ connecting the
sites. The simulations are performed as follows. In an update attempt, we
randomly choose one bond and update the contents of the sites according to
the usual rules: A hole-particle pair within a TASEP is left unchanged,
while a particle-hole pair is always changed to a hole-particle pair. If a
pool-TASEP bond is chosen and the entry site is empty, then a particle is
moved in it with probability $\alpha _{eff,1}$ or $\alpha _{eff,2}$.
Finally, for the TASEP-pool bond, a particle in the last site is moved into
the pool with probability $\beta _{1,2}$. One Monte Carlo step (MCS) is
defined as $L_{1}+L_{2}+2$ attempts.

Starting with $N_{tot}$ particles in the pool (none on the TASEPs), we
allow the system to reach steady-state, which typically takes 100k MCS. For
the next 1M MCS, we record the density profile ($\rho \left( x\right) $) for
each TASEP at every 100 MCS. From these, we compute the overall densities ($%
\rho $), for a total of 10k data points. We also measure the average
currents ($J$), by measuring (for example) the total number of particles
which exit each TASEP over the run and dividing that by $10^{6}$. As in the
earlier study, we are interested in how these quantities are affected by
varying $N_{tot}$. The profiles obviously contain much more detailed
information. Thus, in this first stage, we will mostly report the behavior
of the four functions $\rho _{1,2}\left( N_{tot}\right) $ and $J_{1,2}\left(
N_{tot}\right) $.

Our model has a total of eight parameters: $L_{1}$, $L_{2}$, $\alpha _{1}$, $%
\alpha _{2}$, $\beta _{1}$, $\beta _{2}$, $N_{tot}$, and $N^{\ast }$.To reduce the
number of parameters, we fix $N^{\ast }=1000$. $N^{\ast }$ controls the
strength of the feedback effect for both TASEPs; however, we are focusing on
the effects of having different entry and exit rates, so we will not explore
the effects of $N^{\ast }$ in this study. Since we have different $\alpha $%
's and $\beta $'s, each TASEP can be in a different phase (LD, HD, MC, or
SP) when the pool size becomes large. Thus, 16 different combinations are
possible. From our experience \cite{Adams08,Cook09b}, the most interesting
phenomena occur in the combination HD-HD, the results of which will be
presented next.

\section{Simulation Results}

\label{Section3}

\subsection{HD-HD}

\begin{figure}[htb]
\begin{center}
\includegraphics[width=0.5\textwidth]{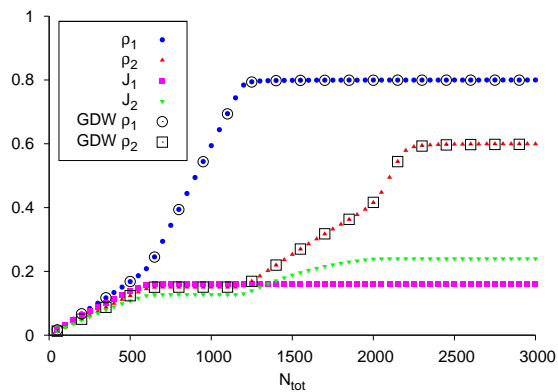}
\caption{Two TASEPs of equal lengths and different rates with $\alpha$'s and $\beta$'s in the HD phase.  The open circles and squares are the results from the domain wall theory presented in section \ref{DW}}
\label{HD-HD-density-all}
\end{center}
\end{figure}

From the earlier study \cite{Adams08}, the overall density of a constrained
HD-TASEP displays three regimes, as $N_{tot}$ is increased: an LD dominated
one, a ``crossover regime'', and one controlled by HD. Respectively, these
are characterized by $\alpha _{eff}\left( N_p\right) <\beta $ , $\alpha
_{eff}\left( N_p\right) =\beta $, and $\alpha _{eff}\left( N_p\right) >\beta 
$. In the crossover regime, $N_p$ remains fixed, while all changes in $%
N_{tot}$ are absorbed by the lattice. Thus, $\rho $ increases {\em linearly}%
, from the LD value of $\beta $ to the HD value of $1-\beta $. The threshold
values of $N_{tot}$ are given by $\alpha _{eff}\left( N_{tot}-\beta L\right)
=\beta $ and $\alpha _{eff}\left( N_{tot}-\left( 1-\beta \right) L\right)
=\beta $. These characteristics are again present when two TASEPs compete
for the pool. The novel features here are the following. If the two TASEPs
make their crossovers at entirely different points, then all changes in $%
N_{tot}$ are absorbed by whichever is in the crossover regime, so that
activity in both the pool and its competitor are completely interrupted. In
Figure \ref{HD-HD-density-all}, we illustrate this phenomenon with the case
of $L_1=1000$, $L_2=1000$, $\alpha _1=0.8$, $\beta _1=0.2$, $\alpha _2=0.6$,
and $\beta _2=0.4$. 
Note first that the two TASEPs fill at different rates
at low $N_{tot}$. This difference is a simple consequence of $\alpha
_{eff,1}\simeq \alpha _1N_p/N^{*}>\alpha _{eff,2}\simeq \alpha _2N_p/N^{*}$.
Next, from $N_{tot}\thicksim 600$ to $\thicksim 1200$, the faster TASEP
makes its crossover while the numbers in the pool and the slower TASEP
remains constant. Thereafter, the slower TASEP continues on its LD regime
and, lastly, makes its crossover in, approximately, the interval $\left[
2000,2200\right] $. We emphasize that, in the respective crossover regimes, $%
\rho _1\in \left[ \beta _1,1-\beta _1\right] $ and $\rho _2\in \left[ \beta
_2,1-\beta _2\right] $.

To understand this effect, we examine the density profile. Even after the
faster TASEP reaches the HD state, its entry rate continues to increase.
This increase results in the decay of the tail near the entrance to changing
as $N_{tot}$ increases, similar to changing $\alpha$ (with fixed $\beta$) in the unconstrained, ordinary TASEP \cite{Derrida92, Derrida93, Schutz93}. As the slower TASEP moves through a crossover
regime, the tail in the profile of the faster TASEP does not change. During
each crossover from LD to HD, the average number of particles in the pool
remains constant. Since each $\alpha _{eff}$ depends on $N_{p}$, the $\alpha
_{eff}$'s also remain constant as $N_{tot}$ increases. The extra particles
from the increase in $N_{tot}$ are added to the TASEP crossing the phase
boundary between the LD and HD phases, resulting in the formation of a
localized shock. A similar phenomenon is found in a single constrained TASEP 
\cite{Adams08} and multiple TASEPs with the same $\alpha $ and $\beta $ \cite{Cook09}.

\begin{figure}[htb]
\begin{center}
\subfigure[]{\includegraphics[width=0.4\textwidth]{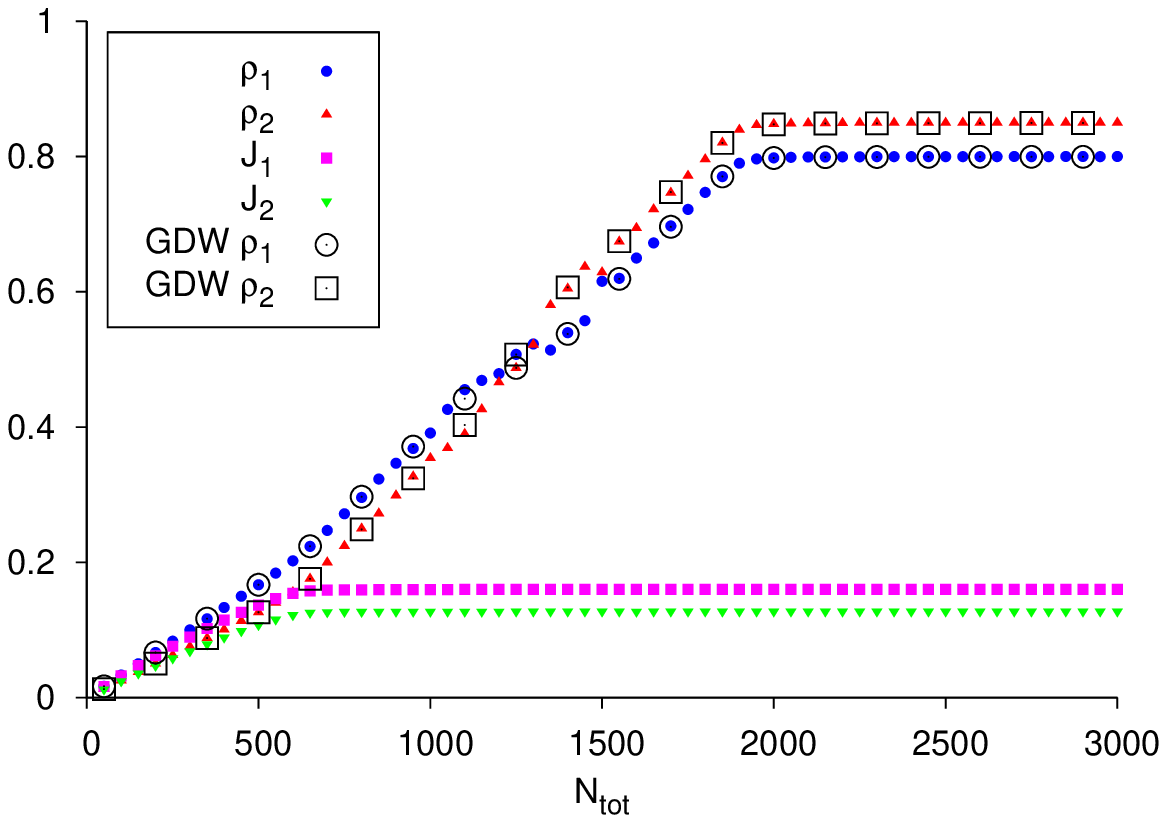}\label{HD-HD-crossover-density-small}}
\subfigure[]{\includegraphics[width=0.4\textwidth]{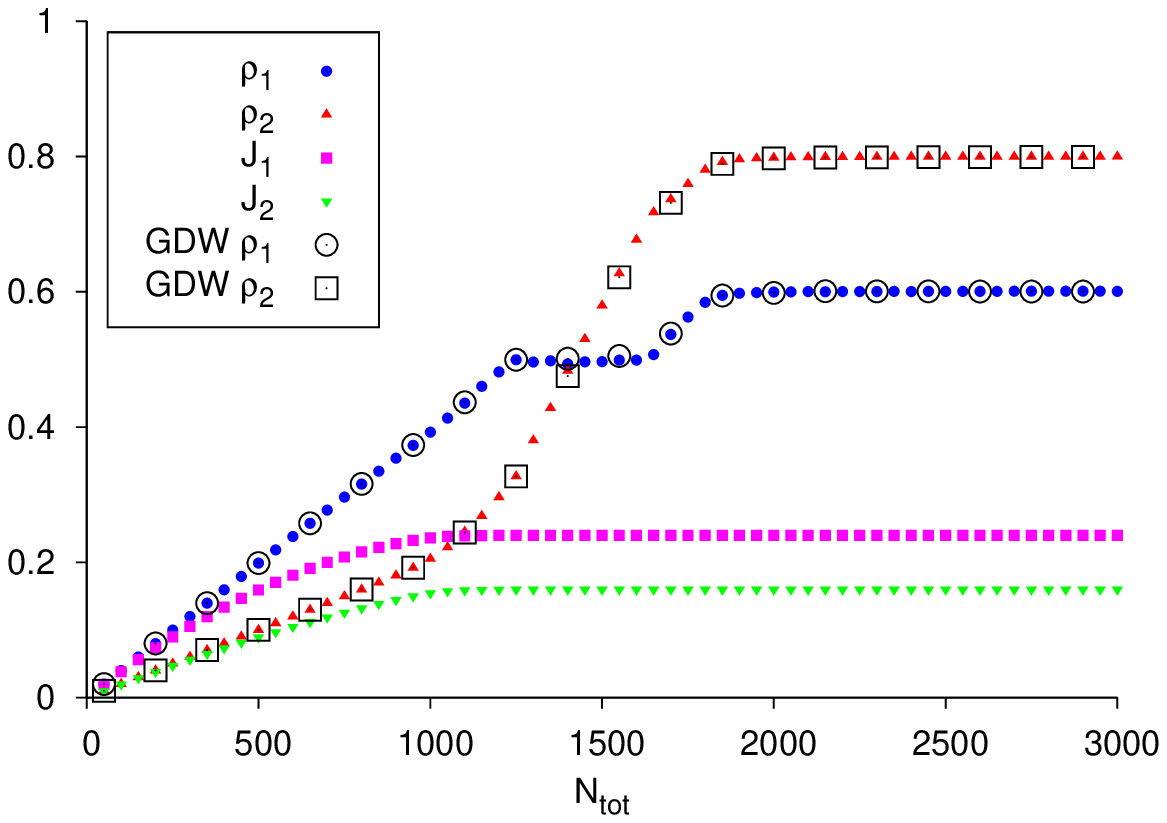}\label{HD-HD-crossover-density-large}}
\caption{Overall densities and currents of two TASEPs of equal lengths entering the crossover regime simultaneously with \subref{HD-HD-crossover-density-small} $\alpha_1/\alpha_2=1.33$ and \subref{HD-HD-crossover-density-large} $\alpha_1/\alpha_2=2$.}
\label{HD-HD-crossover-density}
\end{center}
\end{figure}

For both TASEPs to be in the crossover regime simultaneously, each $\alpha_{eff}$ must reach $\beta$ at the same $N_{tot}$ value.  This condition is achieved when $\alpha_1/\beta_1=\alpha_2/\beta_2$.  The slower TASEP's overall density increases linearly with $N_{tot}$ in the crossover regime, but the faster TASEP's density does not.  Two examples are shown in figure \ref{HD-HD-crossover-density} for $L_1=L_2=1000$, \subref{HD-HD-crossover-density-small} $\alpha_1=0.8$, $\beta_1=0.2$, $\alpha_2=0.6$, $\beta_2=0.15$ and \subref{HD-HD-crossover-density-large} $\alpha_1=1.0$, $\beta_1=0.4$, $\alpha_2=0.5$, $\beta_2=0.2$.
The ratio of $\alpha_1/\alpha_2$ controls the formation of a plateau region for the faster TASEP with a density of $\rho=0.5$.  In our previous study \cite{Cook09b}, similar plateau regions form when the lengths of the TASEPs differed.

\begin{figure}[htb]
\begin{center}
\includegraphics[width=0.5\textwidth]{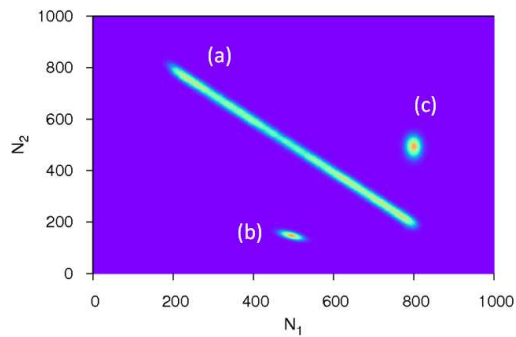}
\caption{Distributions of particle occupation for two TASEPs when (a) both TASEPs are in the crossover regime, (b) one TASEP in the crossover regime and the other in a LD state, and (c) one TASEP in the crossover regime and the other in a HD state.}
\label{HD-HD-dist}
\end{center}
\end{figure}

Another way to visualize the difference of having both TASEPs in the crossover regime is by looking at the probability $P(N_1, N_2)$ to find the $N_1$ particles on the first TASEP and $N_2$ particles on the second.  When $\alpha_1/\beta_1\ne\alpha_2/\beta_2$ and in the crossover regime, the distribution is sharply peaked about the average $N_1$ and $N_2$; otherwise, it is spread across a range of particle occupation pairs whose sum is constant.  These cases are shown in figure \ref{HD-HD-dist} for (a) $L_1=L_2=1000$, $\alpha_1=0.8$, $\beta_1=0.2$, $\alpha_2=0.6$, and $\beta_2=0.15$ with $N_{tot}=1250$ and $L_1=L_2=1000$, $\alpha_1=0.8$, $\beta_1=0.2$, $\alpha_2=0.6$, and $\beta_2=0.40$ with (b) $N_{tot}=900$ and (c) $N_{tot}=2100$.  
The increase of the spread of the distribution when both TASEPs are in the crossover regime comes from the additional degree of freedom that the second TASEP provides in keeping the average number of particles in the pool constant \cite{Cook09b}.  It is important to note that the ranges of $N$ values are the same for both TASEPs and are governed by the exit rate of the faster TASEP, $N_{1,2}\in[\beta L, (1-\beta)L]$.

\begin{figure}[htb]
\begin{center}
\subfigure[]{\includegraphics[width=0.4\textwidth]{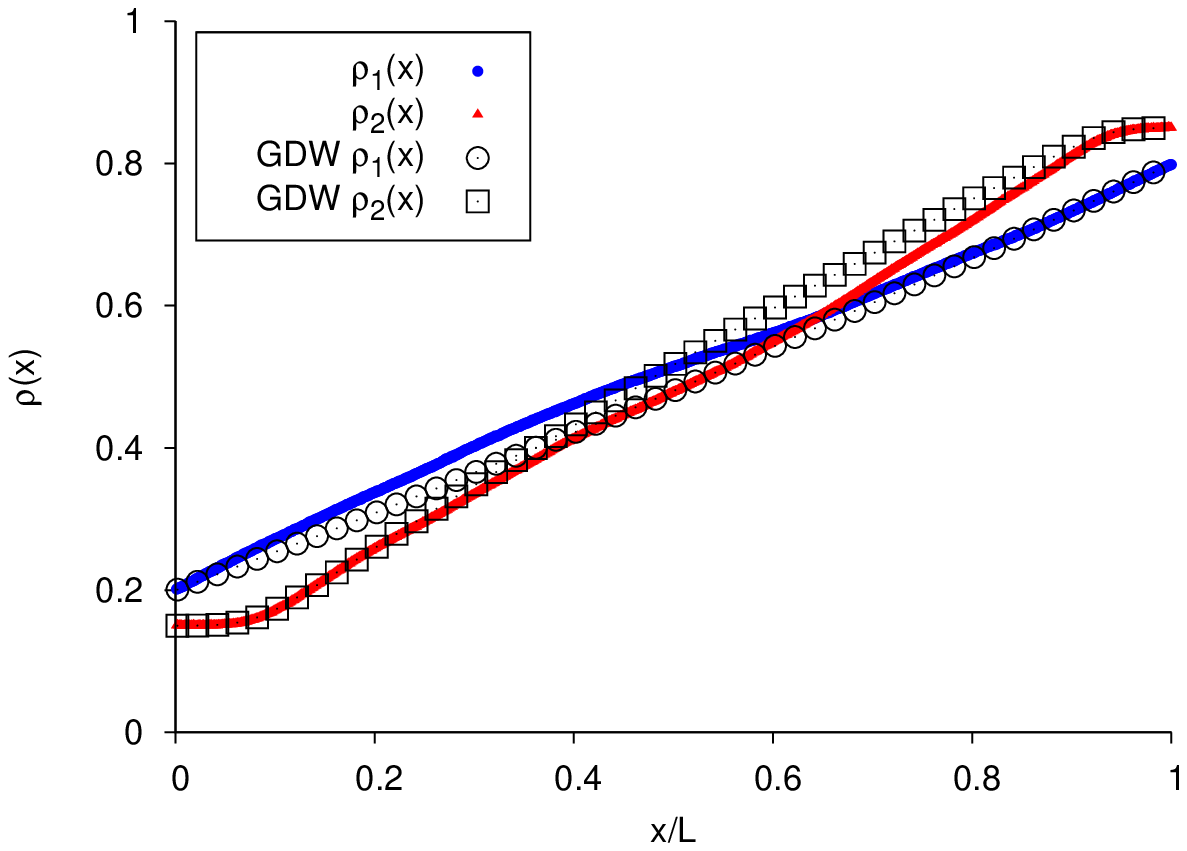}\label{HD-HD-crossover-profile-small}}
\subfigure[]{\includegraphics[width=0.4\textwidth]{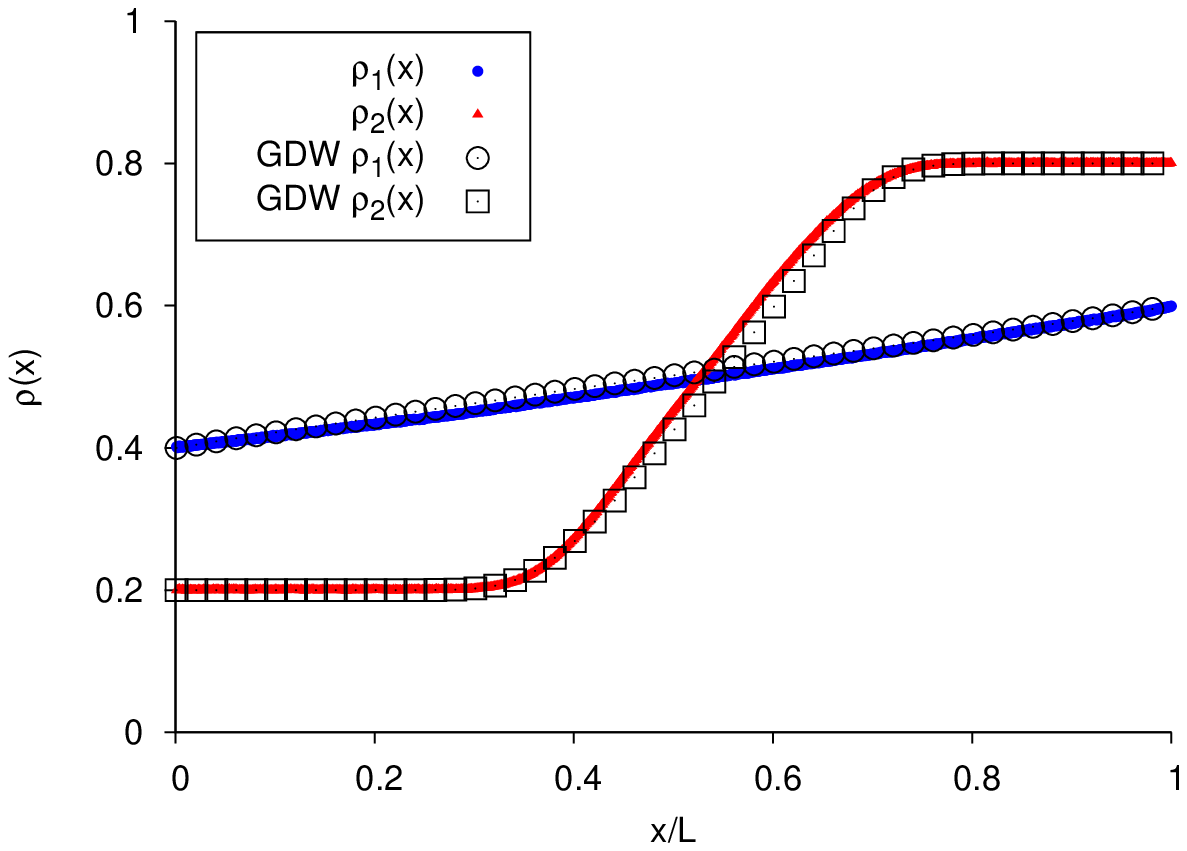}\label{HD-HD-crossover-profile-large}}
\caption{Density profiles of two TASEPs of equal lengths entering the crossover regime simultaneously for \subref{HD-HD-crossover-profile-small} $\alpha_1/\alpha_2=1.33$ with $N_{tot}=1250$ and \subref{HD-HD-crossover-profile-large} $\alpha_1/\alpha_2=2$ with $N_{tot}=1400$.}
\label{HD-HD-crossover-profile}
\end{center}
\end{figure}

To further investigate this crossover regime, we turn to the density profile.  Here, we find that the confinement of the shock between the LD and HD regions is controlled by the ratio of $\alpha_1/\alpha_2$.  Figure \ref{HD-HD-crossover-profile} shows the density profiles for the same set of parameters shown in figure \ref{HD-HD-crossover-density} at \subref{HD-HD-crossover-profile-small} $N_{tot}=1250$ and \subref{HD-HD-crossover-profile-large} $N_{tot}=1400$.
For the simplest TASEP in the SP, the shock performs a random walk over the entire lattice, which results in a linear density profile \cite{Derrida92, Derrida93, Schutz93}.  The linearly increasing regions in the profiles in figure \ref{HD-HD-crossover-profile} indicate the allowed portions of the lattice on which each shock performs a random walk.  The flat regions (of LD or HD) are areas in which the shock does not travel.

Since the number of particles in the pool remain relatively constant in the crossover regime, excess particles are free to choose which lattice to occupy.  Due to the constraint of $\alpha_1/\beta_1=\alpha_2/\beta_2$, the $\alpha_1/\alpha_2$ ratio correlates with the difference between the two shock heights (i.e. the difference between the LD and HD regions densities).  The faster TASEP will always have a smaller shock height, which limits the range of the number of particles it can hold, $[\beta L, (1-\beta)L]$.  The same particle limit applies the the slower TASEP as shown in figure \ref{HD-HD-dist}.  But due to larger shock height, the shock is now confined to a smaller portion of the lattice than the faster TASEP's shock in order to have the same range of $N$ values (thereby keeping the pool size relatively constant).  By decreasing the shock height in the faster TASEP, the range of particles it can hold decreases.  Thus, the shock becomes confined over a smaller region on the slower TASEP as seen in figure \ref{HD-HD-crossover-profile}. This effect was not seen in our previous study \cite{Cook09b} since it is a result of having different entry and exit rates.

\begin{figure}[htb]
\begin{center}
\includegraphics[width=0.5\textwidth]{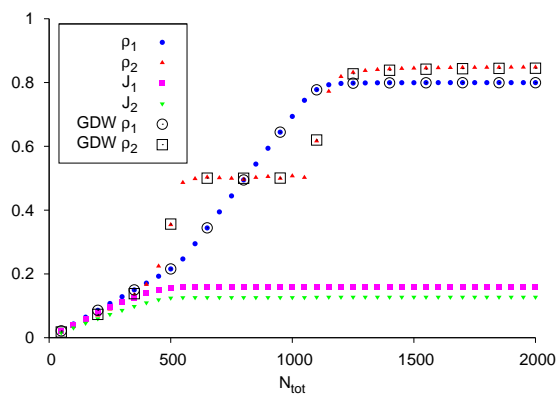}
\caption{Two TASEPs of unequal lengths with different $\alpha$'s and $\beta$'s entering the crossover regime at the same $N_{tot}$ value.}
\label{HD-HD-diffL-density}
\end{center}
\end{figure}

\begin{figure}[htb]
\begin{center}
\includegraphics[width=0.5\textwidth]{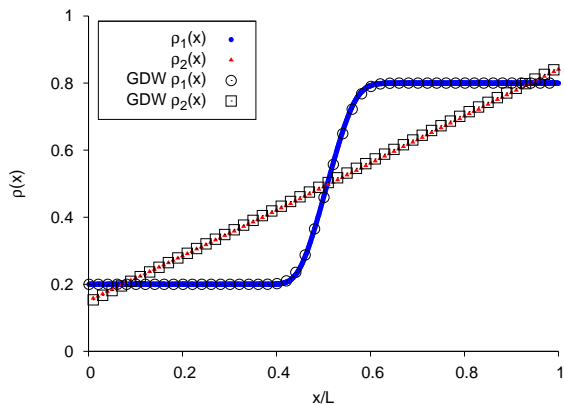}
\caption{Density profiles for two TASEPs with different lengths.}
\label{HD-HD-diffL-profile}
\end{center}
\end{figure}

When both TASEPs enter the crossover regime at the same $N_{tot}$ and the lengths are unequal, we see a trend in the overall density similar to the case of equal rates in \cite{Cook09b}, shown in figure \ref{HD-HD-diffL-density} for $L_1=1000$, $L_2=100$, $\alpha_1=0.8$, $\beta_1=0.2$, $\alpha_2=0.6$, and $\beta_2=0.15$.  
The smaller TASEP has a density of 0.5 for most of the crossover regime, quickly rising to this value from the LD state and from this value to the HD state.  The density profile for this TASEP is linear indicating a delocalized shock.  Figure \ref{HD-HD-diffL-profile} shows this delocalization for $L_1=1000$, $L_2=100$, $\alpha_1=0.8$, $\beta_1=0.2$, $\alpha_2=0.6$, $\beta_2=0.15$, and $N_{tot}=800$.  
The larger TASEP has a localized shock during the crossover regime as seen in the density profile in figure \ref{HD-HD-diffL-profile}.  Even when the rates are reversed, the smaller TASEP has a delocalized shock.  We can conclude that, as long as the size of the smaller TASEP is less than the intrinsic width of the shock localization, the smaller TASEP will have a delocalized shock.

\subsection{HD-SP, HD-LD, and HD-MC}

\begin{figure}[htb]
\begin{center}
\subfigure[]{\includegraphics[width=0.4\textwidth]{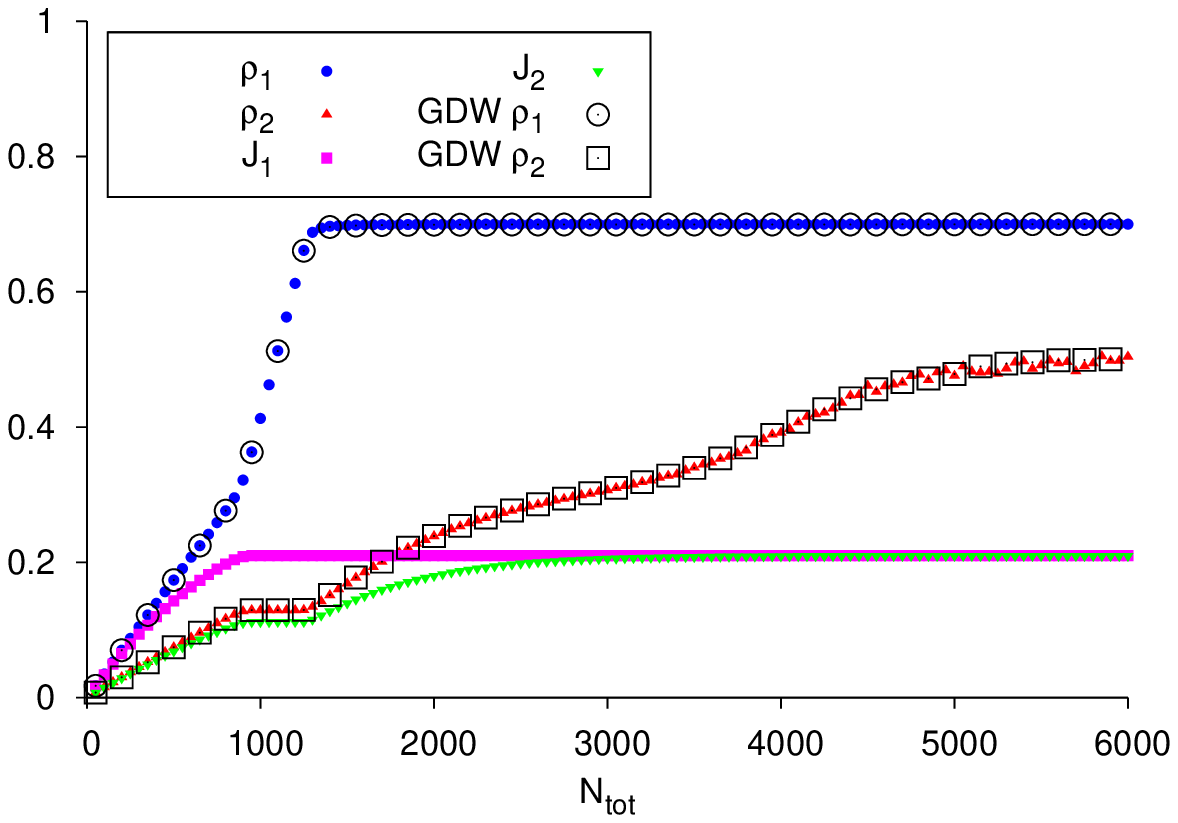}\label{HD-SP-density-equal}}
\subfigure[]{\includegraphics[width=0.4\textwidth]{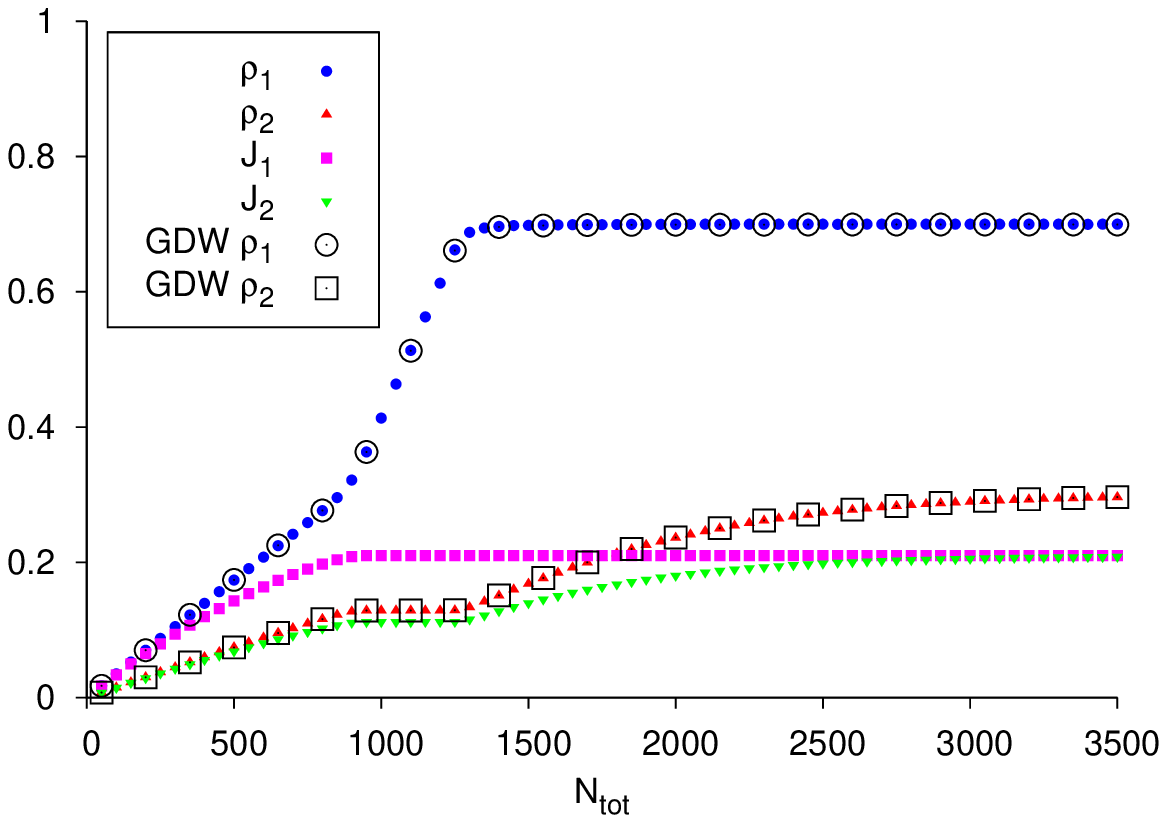}\label{HD-LD-density-equal}}
\subfigure[]{\includegraphics[width=0.4\textwidth]{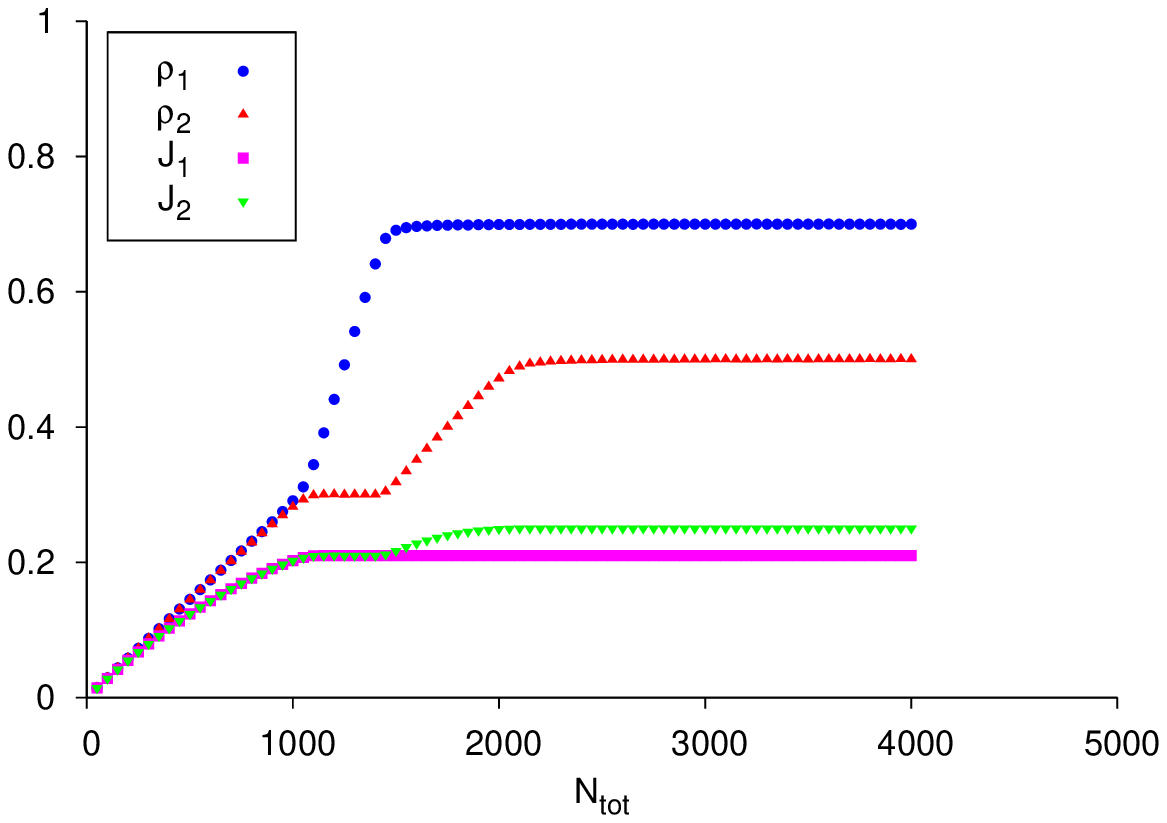}\label{HD-MC-density-equal}}
\caption{Two TASEPs with one in the HD state and the other in the \subref{HD-SP-density-equal} SP, \subref{HD-LD-density-equal} LD, \subref{HD-MC-density-equal} MC state.}
\label{HD-other}
\end{center}
\end{figure}

The combination of having $\alpha$ and $\beta$ on one TASEP in a HD phase with $\alpha$ and $\beta$ on the other TASEP in another phase produces an effect on the density and current similar to having different ratios of $\alpha/\beta$ for each TASEP.  Initially, both TASEPs are in the LD state when $N_{tot}$ is small.  As we increase the number of particles in the system, the HD TASEP begins to crossover from the LD state to a HD one, while the other TASEP's density and current remain constant during this regime.  After the HD TASEP enters its HD state, the other TASEP's density and current continue to increase until it reaches its final state.  Examples of this effect are shown in figures \ref{HD-SP-density-equal}, \ref{HD-LD-density-equal}, and \ref{HD-MC-density-equal} for $L_1=L_2=1000$, $\alpha_1=0.7$, $\beta_1=0.3$ and $\alpha_2=\beta_2=0.3$, $\alpha_2=1-\beta_2=0.3$, $\alpha_2=\beta_2=0.7$, respectively.  
We see no new phenomena when we have different lengths for the TASEPs.  Further, as figure \ref{HD-other} shows, an appropriately generalized domain wall (GDW) theory, which is presented in section \ref{DW}, is quite adequate in predicting the behavior of the overall density (and therefore the current as well).

Finally, we present the data for the HD-MC combination (figure \ref{HD-MC-density-equal}). The various regimes here are easy to understand qualitatively. Since $\alpha_1=\alpha_2$, the initial rise of the densities are the same.  Thereafter, if there were no competition, the behavior of the second 
TASEP would rise smoothly until $\alpha_{2,eff}$ reaches 0.5. But this rise 
is interrupted by the first TASEP traversing the crossover regime, i.e., in the second region.  In 
the third region, it continues its increase and, in the last section, it remains in the MC phase. As the first TASEP has essentially dropped out of the competition, there is no ``kink'' in the transition between these regimes, of course (as better displayed by the current around $N_{tot}\sim2000$).

\subsection{Other phase combinations}

When the $\alpha$'s and $\beta$'s are such that neither TASEP will enter the HD phase, we find the density and current behaving in a manner similar to a single, constrained TASEP \cite{Adams08} as the pool size is increased.  While the total number of particles needed to saturate the system is larger than the number needed for a single TASEP, we find no new features emerging in the overall density and current as a function of $N_{tot}$, even for different lengths.  Some typical results are shown in the figures found in \ref{appendix}.

\section{Theoretical considerations}\label{Section4}

While we presented some qualitative analysis in the previous section, we now supplement those results with a more quantitative analysis for the various phase regimes.

\subsection{LD state}

Regardless of the entry and exit rates, both TASEPs are in a LD state when $N_{tot}$ is small when compared to the smallest lattice length.  From the ordinary TASEP \cite{Derrida93, Schutz01, Evans07}, we know that the overall density is given by $\rho=\alpha$; and for a single TASEP with finite resources \cite{Adams08} it is equal to the average effective entry rate, $\rho=\bar{\alpha}_{eff}$.  Extending these results for two TASEPs with unequal entry and exit rates, we have
\begin{eqnarray}
\rho_1&=\bar{\alpha}_{eff,1}=\alpha_1 f(N_{tot}-\rho_1 L_1-\rho_2 L_2)\\
\rho_2&=\bar{\alpha}_{eff,2}=\alpha_2 f(N_{tot}-\rho_1 L_1-\rho_2 L_2)
\end{eqnarray}
The $\alpha_{eff}$'s depend on both $\rho_1$ and $\rho_2$; therefore, a self-consistent solution is found using these two equations.  For the $f$ chosen in this paper, the solution is found numerically.  Once one of the TASEPs has left the LD state, we must modify our equations.

\subsection{MC state}

If one of the TASEPs enters the MC state, an increase in $\alpha_{eff}$ no longer has an effect on the density or current.  The transition occurs when its $\alpha_{eff}=1/2$.  As with the ordinary TASEP \cite{Derrida92, Derrida93, Schutz93}, the density $\rho=1/2$ and the current $J=1/4$.  The $N_{tot}$ at which this TASEP reaches is final density (assuming $\rho_1$ is entering the MC state) is
\begin{equation}
N_{tot}=f^{-1}\left(\frac{1}{2\alpha_1}\right)+\frac{L_1}{2}+\rho_2 L_2\label{MC-Ntot}
\end{equation}
where $\rho_2$ is the density of the second TASEP.  This density depends on its state,
\begin{equation}
\rho_2=\left\{
\begin{array}{cc}
\frac{\alpha_2}{2\alpha_1} & \rm{LD}\\
\frac{1}{2} & \rm{MC}\\
1-\beta_2 & \rm{HD}
\end{array}
\right. \label{MC-rho2}
\end{equation}
For the parameters shown in figures A\ref{MC-SP-density-equal} and A\ref{LD-MC-density-equal} (where the other TASEP is in the LD state), the MC state is reached at $N_{tot}\sim 1600$.  In figure \ref{HD-MC-density-equal}, the MC state is reached at $N_{tot}\sim 2100$ with the other TASEP in the HD state.  Both TASEPs are approaching the MC state as $N_{tot}$ increases in figure A\ref{MC-MC-density-equal}, where the first one reaches its final density at $N_{tot}\sim 1600$ and the second one at $N_{tot}\sim 2200$.  These values agree with the values predicted by equations \eref{MC-Ntot} and \eref{MC-rho2}.

\subsection{HD crossover}

The HD TASEP enters the crossover regime as $N_{tot}$ increases when $\bar{\alpha}_{eff}=\beta=\rho$ and leaves when $\rho=1-\beta$ \cite{Adams08}.  Taking the HD TASEP to be $\rho_1$, the beginning $N_{tot,1}$ value of the crossover is
\begin{equation}
N_{tot,1}=f^{-1}\left(\frac{\beta_1}{\alpha_1}\right)+\beta_1 L_1+\rho_2 L_2 \label{HD-Ntot1}
\end{equation}
where the value of $\rho_2$ depends on the state of the second TASEP.  Two possibilities exist:  the second TASEP is in either the LD state or HD state.  Then $\rho_2$  is given by
\begin{equation}
\rho_2=\left\{
\begin{array}{cc}
\alpha_2\frac{\beta_1}{\alpha_1} & \rm{LD}\\
1-\beta_2 & \rm{HD}
\end{array}
\right. \label{HD-rho2}
\end{equation}
For the parameters shown in figures \ref{HD-SP-density-equal} and \ref{HD-LD-density-equal} with the second TASEP in the LD state, equations \eref{HD-Ntot1} and \eref{HD-rho2} give a value of $N_{tot,1}\simeq 887$, which agrees with the simulation results.  Using the parameters for figure \ref{HD-MC-density-equal}, we obtain $N_{tot,1}\simeq 1058$, which also agrees with the data shown.  We have similar agreement for figure \ref{HD-HD-density-all} with $N_{tot,1}\simeq 705$ and $N_{tot,1}\simeq 1905$ for each TASEP.

When leaving the crossover regime, the $N_{tot}$ value is given by
\begin{equation}
N_{tot,2}=f^{-1}\left(\frac{\beta_1}{\alpha_1}\right)+(1-\beta_1) L_1+\rho_2 L_2\label{HD-Ntot2}
\end{equation}
where $\rho_2$ is given above.  Equations \eref{HD-Ntot2} and \eref{HD-rho2} give $N_{tot,2}\simeq 1286$ and $N_{tot,2}\simeq 1458$ for the parameters shown in figures \ref{HD-SP-density-equal} and \ref{HD-MC-density-equal}, respectively.  For the parameters shown in figure \ref{HD-HD-density-all}, equations \eref{HD-Ntot2} and \eref{HD-rho2} result in $N_{tot,2}\simeq 1205$ and $N_{tot,2}\simeq 2204$ for each TASEP.  All these values agree with the simulation results.

Determining the density of each TASEP during the crossover regime is simple when only one of the TASEPs is in this regime.  The density of the one in the crossover regime rises linearly with $N_{tot}$, similar to a single constrained TASEP \cite{Adams08}, while the other density remains constant.  Taking $\rho_1$ to be crossing over, we have
\begin{equation}
\rho_1 L_1=N_{tot}-f^{-1}\left(\frac{\beta_1}{\alpha_1}\right)+\rho_2 L_2
\end{equation}
where $\rho_2$ is in either the LD or HD state as before.  However, this simple approach does not work if both TASEPs enter the crossover regime at the same time.  To understand how the density varies with $N_{tot}$ in that situation and with the SP case, we turn to a domain wall approach.

\subsection{Domain wall theory}\label{DW}

The phenomenological domain wall theory has been successfully applied to the unconstrained TASEP \cite{Kolomeisky98, Belitsky02, Santen02} as well as ones with finite resources \cite{Cook09, Cook09b} to understand the steady-state results.  The theory assumes the presence of a sharp domain wall, or shock, separating a low density region near the entrance of a TASEP and a high density region near the exit.  The shock's movement on the lattice depends on the currents of the particles(holes) entering from the entrance(exit) and the wall height \cite{Santen02}.

For an ordinary TASEP with no feedback mechanism, the domain wall moves to the left and to the right with fixed rates that depend on $\alpha$ and $\beta$ \cite{Santen02}.  We generalize this result to include the feedback effect of $\alpha_{eff}$; thus, the hopping rates become site dependent.  While we cannot use the generalized domain wall (GDW) theory when either TASEP is in the MC state, we apply it to all other cases here.  Due to the connection between the shock positions $k_1$, $k_2$, and $N_p$,
\begin{equation}
N_p=N_{tot}-(1-\beta_1)(L_1-k_1)-\alpha_1 f(N_p) k_1-(1-\beta_2)(L_2-k_2)-\alpha_2 f(N_p) k_2
\end{equation}
the function $f(N_p)$ can be rewritten as $f(k_1,k_2)$.  Then $f$ is found using this self-consistent equation.  As the values of $k_1$ and $k_2$ (and subsequently $f$) change, the current of incoming particles $\alpha_{eff}(1-\alpha_{eff})$ and domain wall height $1-\beta-\alpha_{eff}$ for each TASEP will also change due to their dependence on $f$.  This fluctuation in $f$ will lead to domain wall hopping rates that are site dependent \cite{Cook09, Cook09b}.

With two TASEPs connected to a single finite pool of particles, the probability $P$ of finding a set of domain wall positions $\{k_1,k_2\}$ at steady-state is given by
\begin{eqnarray}
\nonumber\fl 0=D^+_{k_1-1,k_2}P(k_1-1,k_2)+D^-_{k_1+1,k_2}P(k_1+1,k_2)+E^+_{k_1,k_2-1}P(k_1,k_2-1)\\
+E^-_{k_1,k_2+1}P(k_1,k_2+1)-\left(D^+_{k_1,k_2}+D^-_{k_1,k_2}+E^+_{k_1,k_2}+E^-_{k_1,k_2}\right)P(k_1,k_2)
\end{eqnarray}
where
\begin{eqnarray}
D^-_{k_1,k_2}&=\frac{\alpha_1 f(k_1,k_2)(1-\alpha_1 f(k_1,k_2))}{1-\beta_1-\alpha_1 f(k_1,k_2)}\\
D^+_{k_1,k_2}&=\frac{\beta_1(1-\beta_1)}{1-\beta_1-\alpha_1 f(k_1,k_2)}\\
E^-_{k_1,k_2}&=\frac{\alpha_2 f(k_1,k_2)(1-\alpha_2 f(k_1,k_2))}{1-\beta_2-\alpha_2 f(k_1,k_2)}\\
E^+_{k_1,k_2}&=\frac{\beta_1(1-\beta_2)}{1-\beta_2-\alpha_2 f(k_1,k_2)}
\end{eqnarray}
along with appropriate reflecting boundary conditions.  We lose the detailed balance that was previously exploited to find an analytical solution \cite{Cook09b}.  While it is possible to find the $P(k_1,k_2)$ analytically, it is not very practical.  This system of $(L_1+1)(L_2+1)$ equations ($(L_1+1)(L_2+1)-1$ which are linearly independent) becomes time-consuming to solve even for numerically finding the eigenvector corresponding to the zero eigenvalue.  Instead, we build the probability distribution through Monte Carlo simulations of a random walker on a two-dimensional lattice with the hopping rates $D^+$, $D^-$, $E^+$, and $E^-$.  These simulations give us the $P(k_1,k_2)$ we need to calculate the density profile and overall density.  The profile for each TASEP is given by \cite{Cook09b}
\begin{eqnarray}
\rho_1(x)&=\sum_{k_2=0}^{L_2}\left[\sum_{k_1=0}^x (1-\beta_1)P(k_1,k_2)+\sum_{x+1}^{L_1} \alpha_1f(k_1,k_2)P(k_1,k_2)\right]\\
\rho_2(x)&=\sum_{k_1=0}^{L_1}\left[\sum_{k_2=0}^x (1-\beta_1)P(k_1,k_2)+\sum_{x+1}^{L_2} \alpha_1f(k_1,k_2)P(k_1,k_2)\right]
\end{eqnarray}
The overall density is given by $\rho_i=\sum_{x=1}^{L_i}\rho_i(x)$.  The GDW theory results agree with the simulation results as shown in figures \ref{HD-HD-density-all}, \ref{HD-HD-crossover-density}, \ref{HD-HD-diffL-density}, \ref{HD-SP-density-equal}, \ref{HD-LD-density-equal}, A\ref{LD-LD-density-equal}, A\ref{LD-SP-density-equal}, A\ref{SP-SP-density-equal} for the overall density, and figures \ref{HD-HD-crossover-profile}, \ref{HD-HD-diffL-profile} for the density profile.

The domain wall picture helps explain the difference between the results in figure \ref{HD-HD-diffL-density} and \ref{HD-HD-crossover-density-large}.  In figure \ref{HD-HD-diffL-density}, the delocalization of the shock over a range of $N_{tot}$ is due to the domain wall reflecting at the boundaries on the smaller TASEP.  In figure \ref{HD-HD-crossover-density-large}, the difference in hopping rates allow the shock in the faster TASEP (larger rates) to move about the entire lattice more easily than the one on the slower TASEP (smaller rates).  The shock in the slower TASEP will be less likely to move away from its average position, leading to shock localization.  Also, the domain wall height, which appears in the denominator of the hopping rates, plays a significant role.  If the wall height is too large, then the difference between the rates for each TASEP decreases.  The smaller difference allows the shock to wander over a large portion of the slower TASEP, as seen in figure \ref{HD-HD-crossover-profile-small}.  Thus, shock localization can be induced by either different lengths or different rates.

Finally, associated with figure \ref{HD-MC-density-equal} (HD-MC), we have no GDW theory to provide a good theoretical prediction, as the second TASEP ends in a state with no domain walls (MC). Since the general aspects of this competition is qualitatively understood, designing a more sophisticated and quantitative theory seems unnecessary.

\section{Summary and Outlook}\label{Section5}

In this paper, we explored how competition for particles between two TASEPs affect the overall density, density profile, and current.  Through simulation results and theoretical considerations, we have shown that new effects arise from having different entry and exit rates on the TASEPs.  One of these effects is the localization of a shock on the lattice due to the difference in entry and exit rates.  The appropriately generalized domain wall theory captured the shock localization phenomenon and reproduced the overall density and density profiles.  However, more work still needs to be done if we want to make a connection to the translation process in a cell.

While our study has focused on only two TASEPs, more should be added.  Recalling our motivation of protein synthesis, many mRNA's compete for the same pool of ribosomes.  The parameter space to explore increases with each additional TASEP, which could lead to new phenomena occurring.  Similarly, the dimension of the random walk set forth in the GDW theory increases for each new TASEP that is added.  A systematic study of multiple TASEPs would be useful.

Beyond multiple TASEPs, other additions to the model should be made in order to better model the translation process during protein synthesis \cite{MacDonald68,MacDonald69,Shaw03,Chou03,Zia11}. First, the ribosome does not move to the next codon at the same rate for all codons, and the rate may depend on the concentration of amino acid transfer-RNAs (aa-tRNA) in the cell \cite{Dong96}. Thus, TASEPs with inhomogeneous, mRNA-sequence dependent, hopping rates must be taken into account \cite{Shaw03,Zia11}. Now that these rates depend on the aa-tRNA concentrations, it is reasonable to consider the competition for finite aa-tRNA resources. Notably, such an ambitious undertaking has been carried out recently \cite{Brackley10,Brackley11}, although the behavior in a real cell, with thousands of copies of thousands of different genes, will remain difficult for simulation studies in the conceivable future. Second, ribosomes cover more than one codon, typically 12 \cite{cell}. Therefore, the size of the particles should be larger as well \cite{MacDonald68,MacDonald69,Shaw03,Chou03,Dong07}. By combining these individual elements, we hope to gain a better understanding of the translation process during protein synthesis, as well as non-equilibrium systems in general.

After completing this work, we became aware of a similar study by P.\ Greulich, et.\ al.\ \cite{Greulich12}. The main differences between our efforts are the following. 1) We explore the density profile in our Monte Carlo simulations and theoretical approaches. 2) We distinguish between systems in the SP with localized shocks and those with delocalized ones. 3) We explain our results from a domain wall perspective for both the overall density and density profile, instead of using a mean-field approach as in \cite{Greulich12}.

\section*{Acknowledgments}

We would like to thank Jiajia Dong and Beate Schmittmann for insightful discussions, Irina Mazilu and Tom Williams for a critical reading of the manuscript, and Martin Evans for calling our attention to ref. \cite{Greulich12}.  This work was funded in part by the U.S. National Science Foundation through Grant No.\ DMR-1005417, and Washington and Lee University through the Lenfest Grant.

\newpage
\appendix
\section{Results for cases without an HD phase}\label{appendix}
\begin{figure}[htb]
\begin{center}
\subfigure[]{\includegraphics[width=0.35\textwidth]{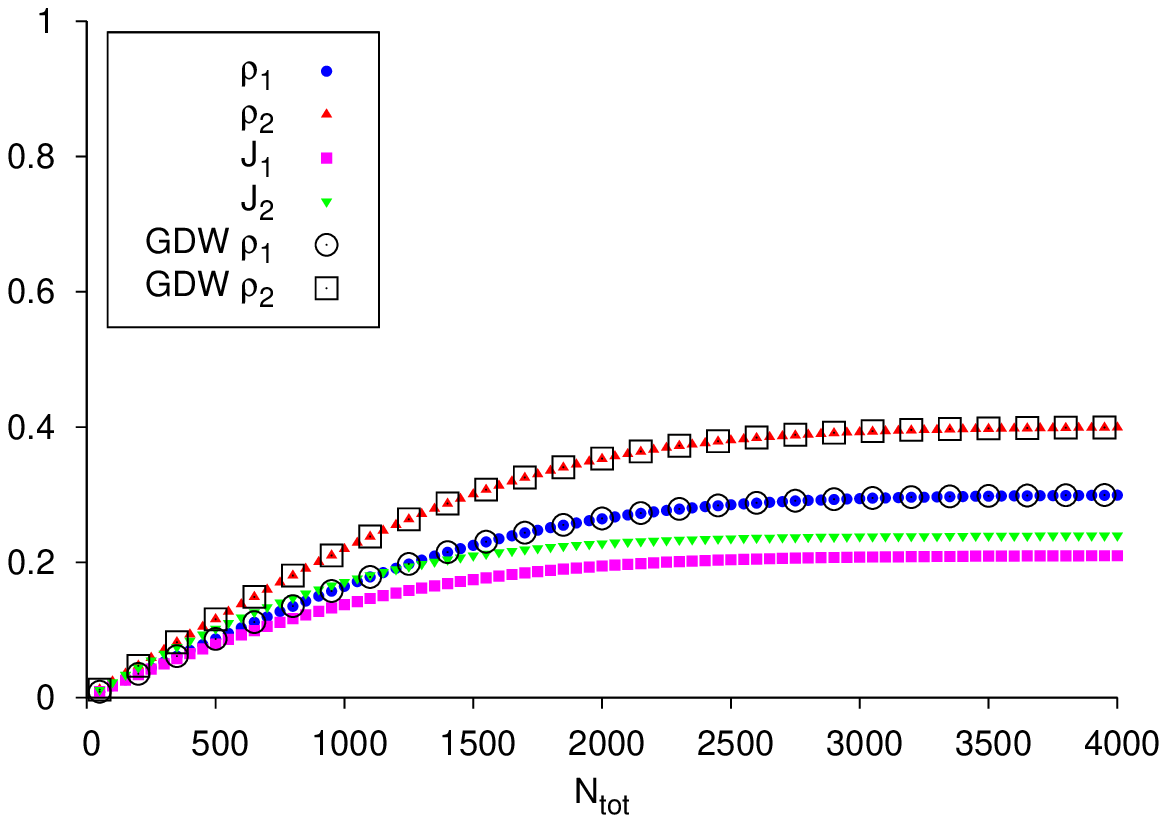}\label{LD-LD-density-equal}}
\subfigure[]{\includegraphics[width=0.35\textwidth]{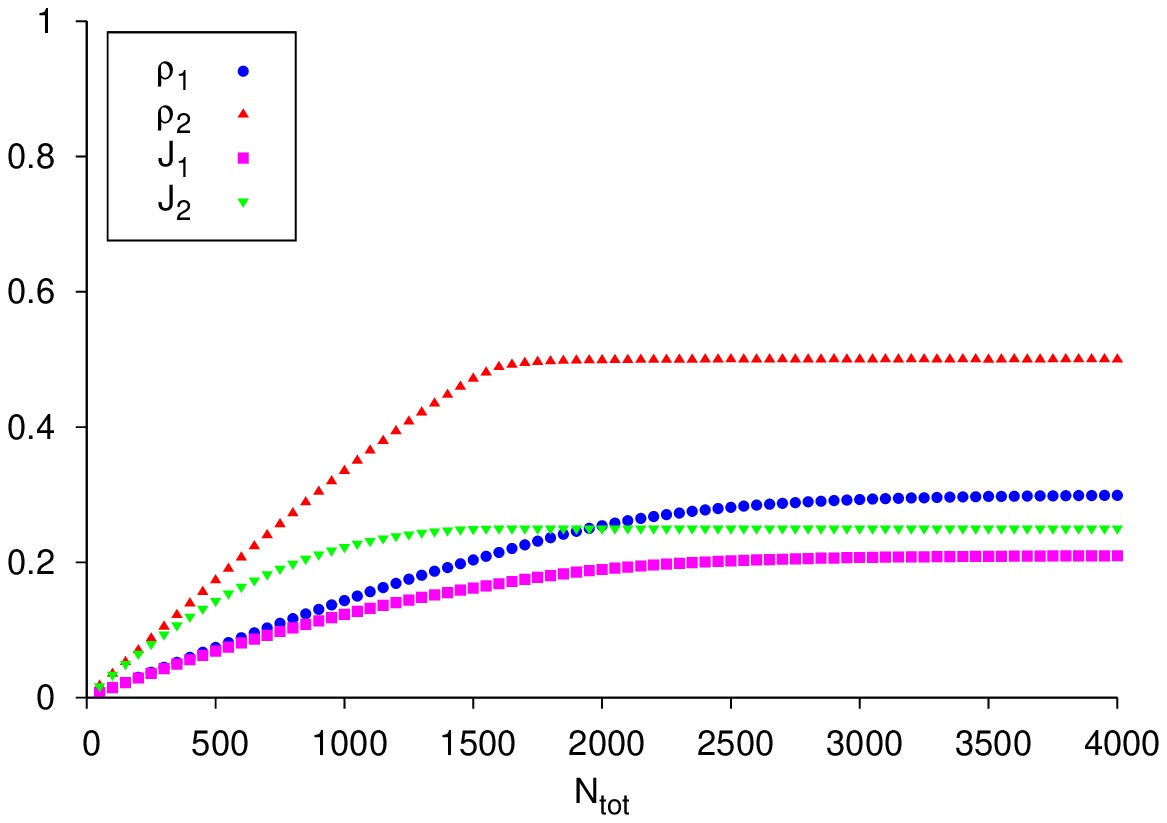}\label{LD-MC-density-equal}}
\subfigure[]{\includegraphics[width=0.35\textwidth]{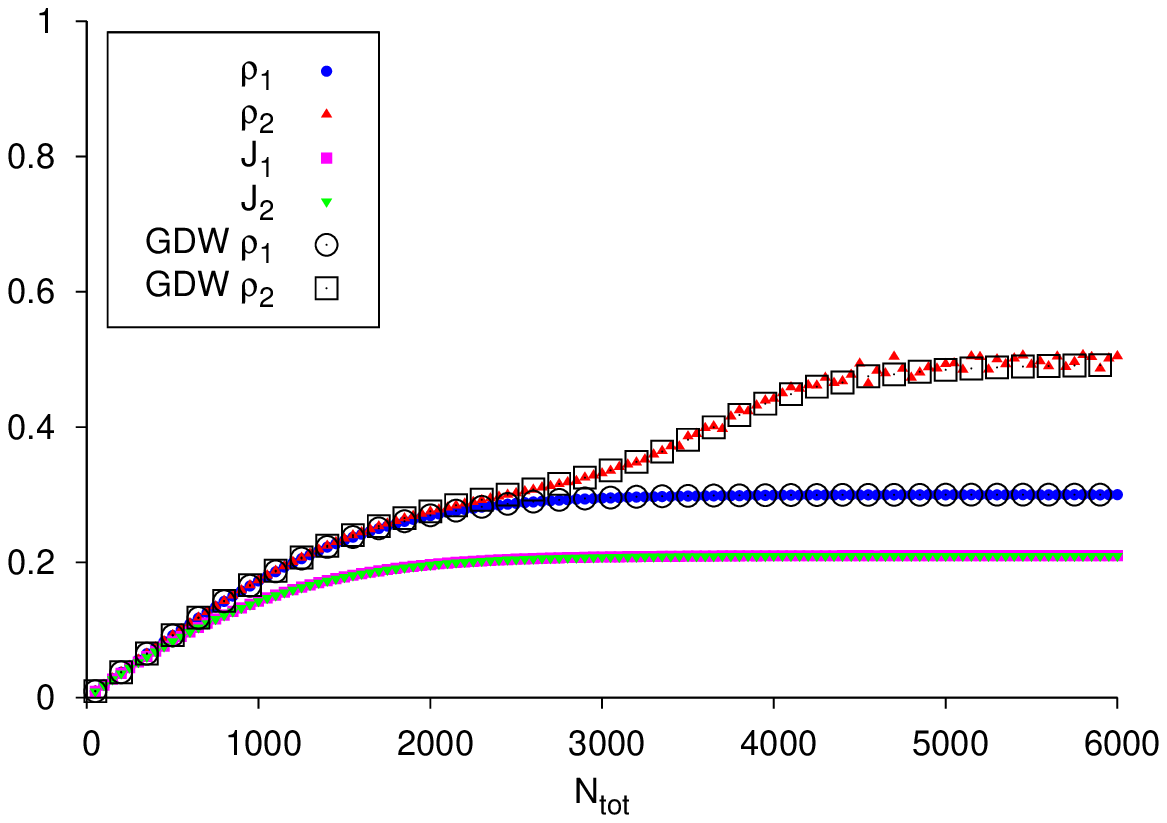}\label{LD-SP-density-equal}}
\subfigure[]{\includegraphics[width=0.35\textwidth]{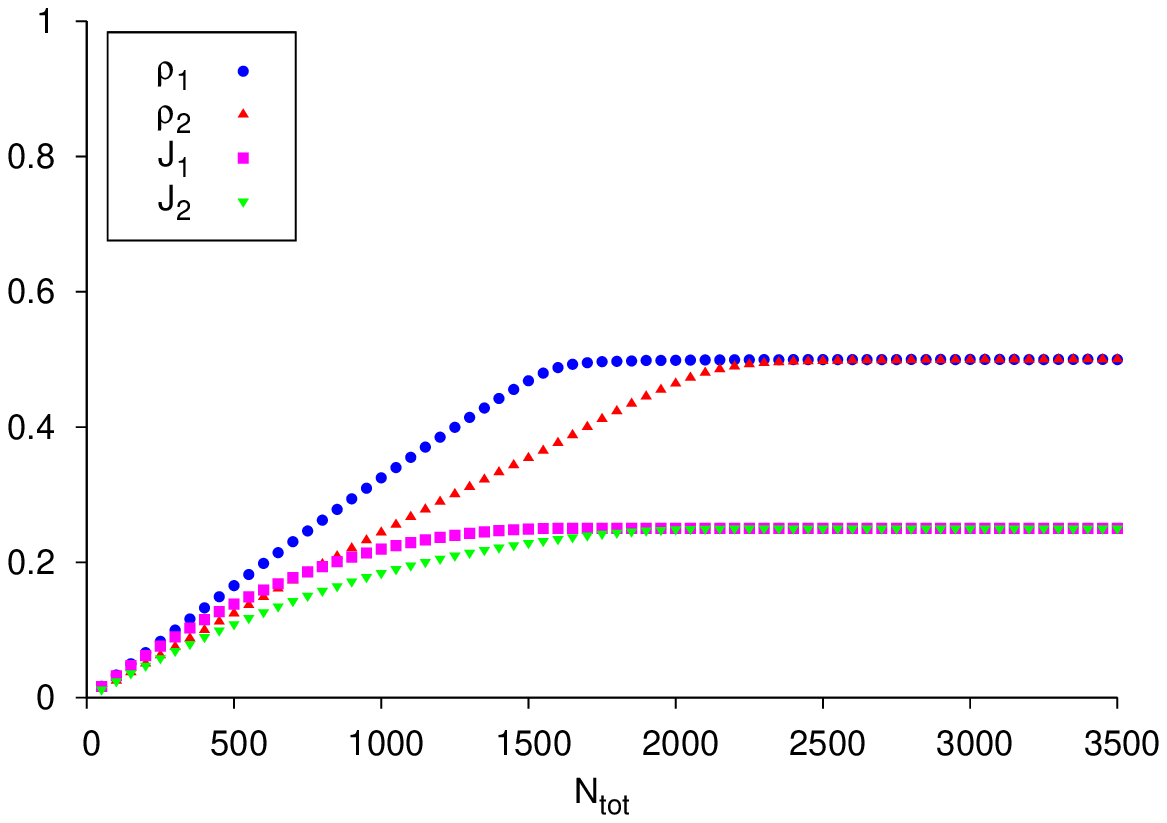}\label{MC-MC-density-equal}}
\subfigure[]{\includegraphics[width=0.35\textwidth]{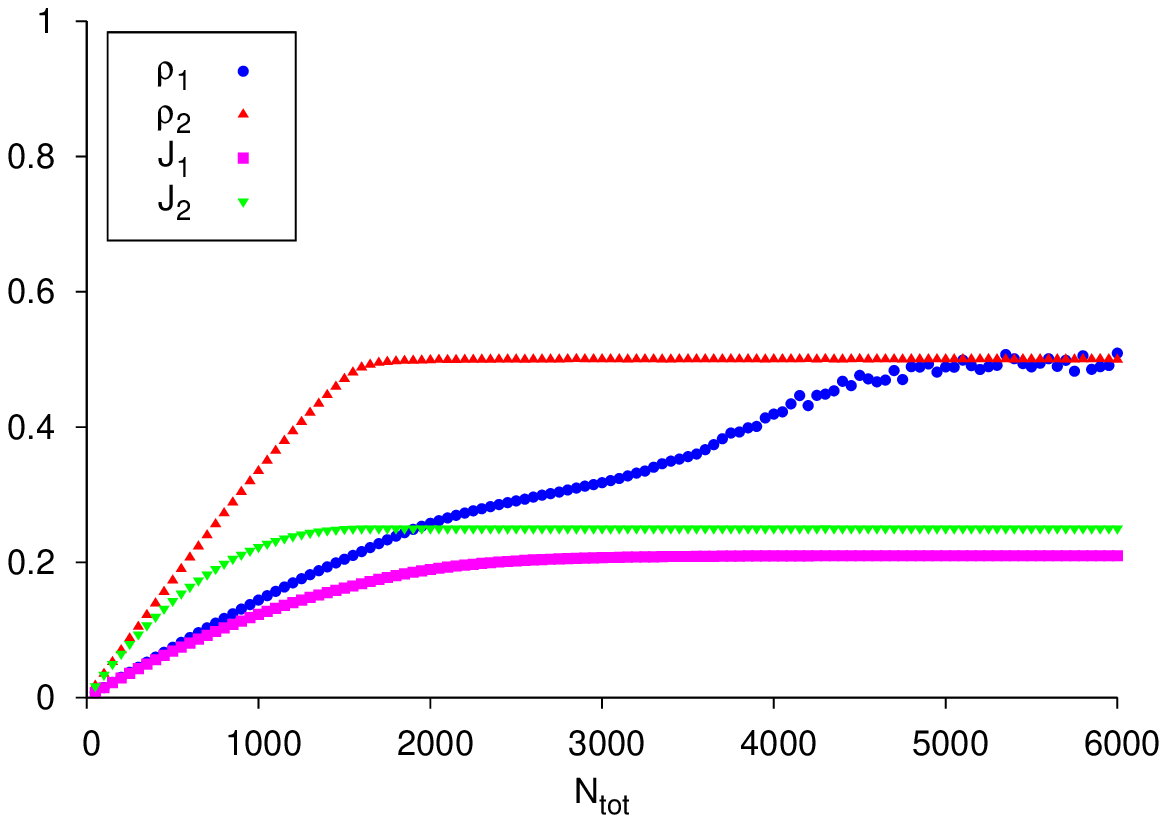}\label{MC-SP-density-equal}}
\subfigure[]{\includegraphics[width=0.35\textwidth]{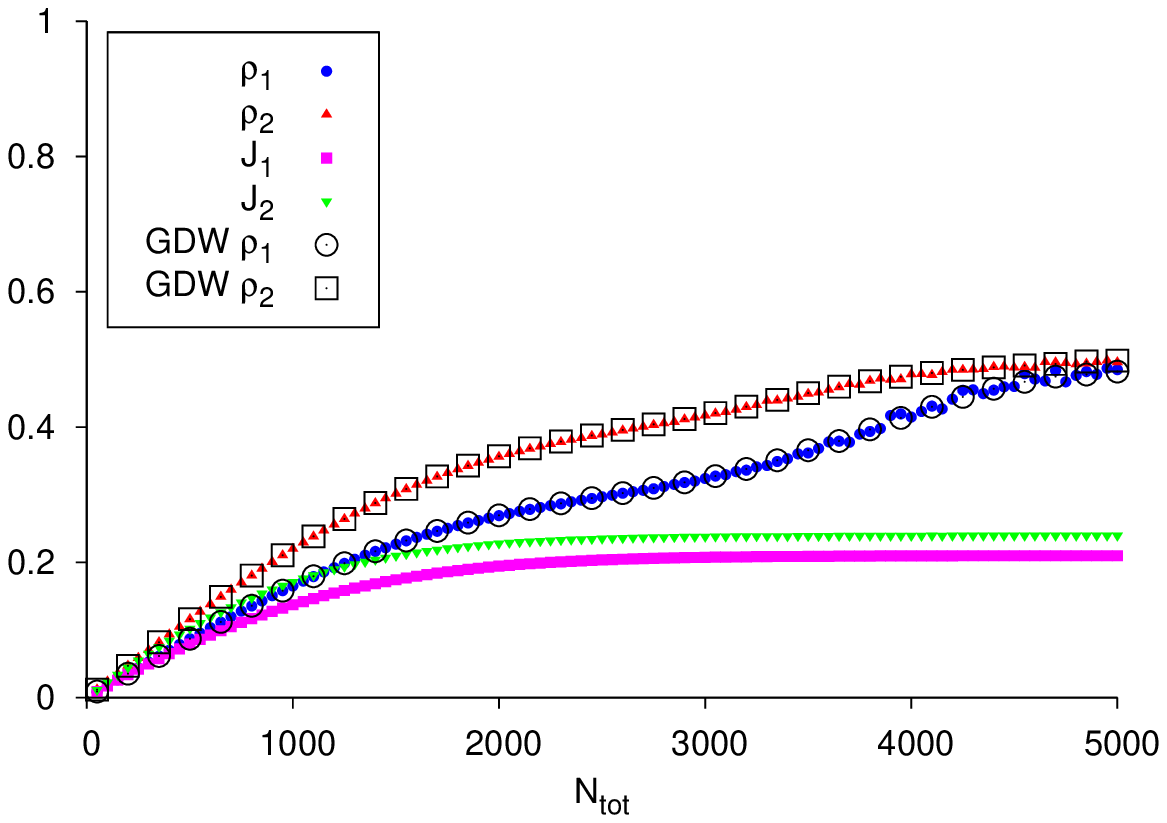}\label{SP-SP-density-equal}}
\caption{Overall density and current for various phase combinations.}
\label{other-phases}
\end{center}
\end{figure}

Results in figure \ref{other-phases} are shown for $L_1=L_2=1000$, $N^*=1000$, and \subref{LD-LD-density-equal} $\alpha_1=1-\beta_1=0.3$, $\alpha_2=1-\beta_2=0.4$; \subref{LD-MC-density-equal} $\alpha_1=1-\beta_1=0.3$, $\alpha_2=\beta_2=0.7$; \subref{LD-SP-density-equal} $\alpha_1=1-\beta_1=0.3$, $\alpha_2=\beta_2=0.3$; \subref{MC-MC-density-equal} $\alpha_1=\beta_1=0.8$, $\alpha_2=\beta_2=0.6$; \subref{MC-SP-density-equal}
$\alpha_1=\beta_1=0.7$, $\alpha_2=\beta_2=0.3$; \subref{SP-SP-density-equal}
$\alpha_1=\beta_1=0.3$, $\alpha_2=\beta_2=0.4$.

\section*{References}
\bibliographystyle{iopart-num} 
\bibliography{references}

\providecommand{\newblock}{}
\begin{thebibliography}{10}
\expandafter\ifx\csname url\endcsname\relax
  \def\url#1{{\tt #1}}\fi
\expandafter\ifx\csname urlprefix\endcsname\relax\def\urlprefix{URL }\fi
\providecommand{\eprint}[2][]{\url{#2}}

\bibitem{CMMP10}
Committee~on CMMP~2010 Solid State Science~Committee N~R~C 2007 {\em
  Condensed-Matter and Materials Physics: The Science of the World Around Us\/}
  (Washington, DC: National Academies Press)

\bibitem{MacDonald68}
MacDonald C~T, Gibbs J~H and Pipkin A~C 1968 {\em Biopolymers\/} {\bf 6}

\bibitem{MacDonald69}
MacDonald C~T and Gibbs J~H 1969 {\em Biopolymers\/} {\bf 7} 707

\bibitem{Shaw03}
Shaw L~B, Zia R~K~P and Lee K~H 2003 {\em Phys. Rev. E\/} {\bf 68} 021910

\bibitem{Chou03}
{Lakatos} G and {Chou} T 2003 {\em J. Phys. A: Math. Gen.\/} {\bf 36} 2027

\bibitem{Chowdhury00}
Chowdhury D, Santen L and Schadschneider A 2000 {\em Phys. Rep.\/} {\bf 329}
  199

\bibitem{Popkov01}
Popkov V, Santen L, Schadschneider A and Sch\"{u}tz G~M 2001 {\em J. Phys. A:
  Math. Gen.\/} {\bf 34} L45

\bibitem{Kardar86}
Kardar M, Parisi G and Zhang Y~C 1986 {\em Phys. Rev. Lett.\/} {\bf 56} 889

\bibitem{Wolf90}
Wolf D~E and Tang L~H 1990 {\em Phys. Rev. Lett.\/} {\bf 65} 1591

\bibitem{Spitzer70}
Spitzer F 1970 {\em Adv. Math.\/} {\bf 5} 246

\bibitem{DeMasi85}
De~Masi A and Ferrari P~A 1985 {\em J. Stat. Phys.\/} {\bf 38} 603

\bibitem{Kutner85}
Kutner R and van Beijeren H 1985 {\em J. Stat. Phys.\/} {\bf 39} 317

\bibitem{Dhar87}
Dhar D 1987 {\em Phase Transit.\/} {\bf 9} 51

\bibitem{Majumdar91}
Majumdar S~N and Barma M 1991 {\em Phys. Rev. B\/} {\bf 44} 5306

\bibitem{Gwa92}
Gwa L~H and Spohn H 1992 {\em Phys. Rev. A\/} {\bf 46} 844

\bibitem{Derrida93b}
Derrida B, Evans M~R and Mukamel D 1993 {\em J. Phys. A: Math. Gen.\/} {\bf 26}
  4911

\bibitem{Kim95}
Kim D 1995 {\em Phys. Rev. E\/} {\bf 52} 3512

\bibitem{Golinelli05}
Golinelli O and Mallick K 2005 {\em J. Phys. A: Math. Gen.\/} {\bf 38} 1419

\bibitem{Krug91}
Krug J 1991 {\em Phys. Rev. Lett.\/} {\bf 67} 1882

\bibitem{Derrida92}
Derrida B, Domany E and Mukamel D 1992 {\em J. Stat. Phys.\/} {\bf 69} 667

\bibitem{Derrida93}
Derrida B, Evans M~R, Hakim V and Pasquier V 1993 {\em J. Phys. A: Math.
  Gen.\/} {\bf 26} 1493

\bibitem{Schutz93}
Sch\"{u}tz G and Domany E 1993 {\em J. Stat. Phys.\/} {\bf 72} 277

\bibitem{Pierobon05}
{Pierobon} P, {Parmeggiani} A, {von Oppen} F and {Frey} E 2005 {\em Phys. Rev.
  E\/} {\bf 72} 036123

\bibitem{Dudzinski00}
Dudzinski M and Sch\"{u}tz G~M 2000 {\em J. Phys. A: Math. Gen.\/} {\bf 33}
  8351

\bibitem{Nagy02}
{Nagy} Z, {Appert} C and {Santen} L 2002 {\em J. Stat. Phys.\/} {\bf 109} 634

\bibitem{Takesue03}
Takesue S, Mitsudo T and Hayakawa H 2003 {\em Phys. Rev. E\/} {\bf 68} 015103

\bibitem{deGier06}
de~Gier J and Essler F~H~L 2006 {\em J. Stat. Mech.\/}  P12011

\bibitem{Gupta07}
Gupta S, Majumdar S~N, Godr\`eche C and Barma M 2007 {\em Phys. Rev. E\/} {\bf
  76} 021112

\bibitem{Chou11}
Chou T, Mallick K and Zia R~K~P 2011 {\em Rep. Prog. Phys.\/} {\bf 74} 116601

\bibitem{Dong07}
{Dong} J~J, {Schmittmann} B and {Zia} R~K~P 2007 {\em J. Stat. Phys.\/} {\bf
  128} 21

\bibitem{Chou04}
Chou T and Lakatos G 2004 {\em Phys. Rev. Lett.\/} {\bf 93} 198101

\bibitem{Dong07b}
{Dong} J~J, {Schmittmann} B and {Zia} R~K~P 2007 {\em Phys. Rev. E\/} {\bf 76}
  051113

\bibitem{Chou03b}
{Chou} T 2003 {\em Biophys. J.\/} {\bf 85} 755

\bibitem{Adams08}
Adams D~A, Schmittmann B and Zia R~K~P 2008 {\em J. Stat. Mech.\/}  P06009

\bibitem{Cook09}
Cook L~J and Zia R~K~P 2009 {\em J. Stat. Mech.\/}  P02012

\bibitem{Cook09b}
Cook L~J, Zia R~K~P and Schmittmann B 2009 {\em Phys. Rev. E\/} {\bf 80} 031142

\bibitem{Schutz01}
Sch\"{u}tz G~M 2001 Exactly solvable models for many-body systems far from
  equilibrium {\em Phase transitions and critical phenomena\/} vol~19 ed Domb C
  and Lebowitz J (Academic Press)

\bibitem{Evans07}
Blythe R~A and Evans M~R 2007 {\em J. Phys. A: Math. Gen.\/} {\bf 40} R333

\bibitem{Kolomeisky98}
Kolomeisky A~B, Sch\"{u}tz G~M, Kolomeisky E~B and Straley J~P 1998 {\em J.
  Phys. A: Math. Gen.\/} {\bf 31} 6911

\bibitem{Belitsky02}
Belitsky V and Sch\"{u}tz G 2002 {\em Electron. J. Probab.\/} {\bf 7} 1

\bibitem{Santen02}
Santen L and Appert C 2002 {\em J. Stat. Phys.\/} {\bf 106} 187

\bibitem{Zia11}
Zia R, Dong J and Schmittmann B 2011 {\em Journal of Statistical Physics\/}
  {\bf 144} 405

\bibitem{Dong96}
Dong H, Nilsson L and Kurland C 1996 {\em Journal of Molecular Biology\/} {\bf
  260} 649

\bibitem{Brackley10}
Brackley C~A, Romano M~C and Thiel M 2010 {\em Phys. Rev. E\/} {\bf 82} 051920

\bibitem{Brackley11}
{Brackley} C~A, {Romano} M~C and {Thiel} M 2011 {\em PLoS Comput. Biol.\/} {\bf
  7} e1002203

\bibitem{cell}
Alberts B, Johnson A, Lewis J, Raff M, Roberts K and Walter P 2007 {\em
  Molecular biology of the cell\/} (New York: Garland Science) ISBN
  978-0-8153-4105-5

\bibitem{Greulich12}
Greulich P, Ciandrini L, Allen R~J and Romano M~C 2012 {\em Phys. Rev. E\/}
  {\bf 85} 011142

\end{thebibliography}

\end{document}